# Ultracompact Field Effect Electro-Absorption Plasmonic Modulator

Kaifeng Shi[*] and Zhaolin Lu[*,†]

[*]*Microsystems Engineering PhD Program*
[†]*Department of Electrical and Microelectronic Engineering*
*Rochester Institute of Technology, Rochester, New York, 14623, USA*
zhaolin.lu@rit.edu

**Abstract**

One of the technical barriers impeding the wide applications of integrated photonic circuits is the lack of ultracompact, high speed, broadband electro-optical (EO) modulators, which up-convert electronic signals into high bit-rate photonic data. In addition to direct modulation of lasers, EO modulators can be classified into (i) phase modulation based on EO effect or free-carrier injection [1, 2], or (ii) absorption modulation based on Franz-Keldysh effect or quantum-confined Stark effect [3- 7]. Due to the poor EO properties of regular materials, a conventional EO modulator has a very large footprint. Based on high-Q resonators, recent efforts have advanced EO modulators [8,9,10] into microscale footprints, which have nearly reached their physical limits restricted by the materials. On-chip optical interconnects require ultrafast EO modulators at the nanoscale. The technical barrier may not be well overcome based on conventional approaches and well-known materials. Herein, we report an EO modulator, more specifically electro-absorption (EA) modulator, based on the integration of a novel yet inexpensive active material, indium tin oxide (ITO), in a metal-insulator-metal (MIM) plasmonic waveguide platform, where the field effect is then greatly enhanced by high-k insulator and double capacitor gating scheme. The modulator waveguide length is only 800 nm, which is the smallest recorded dimension according to our knowledge. Preliminary results show that it has extinction ratio of 1.75 (2.43 dB) at 10 MHz, works up to 500 MHz (limited by testing setup for now), and can potentially operate at high speed.

Most of previous effort was focused on the exploration of the EO properties of dielectrics or polymers owing to their low optical absorption for waveguide applications, whereas the optical properties of absorptive materials, for example conducting oxides (COs), have been relatively overlooked, until recent work showing that the optical dielectric constant of COs in the charged layer of a metal-insulator-CO (MIC) structure can be tuned in a large range by electrical gating [11]. When a large electric field is applied across the insulator layer of an MIC structure, significant surface charge can be induced and the induced charge can greatly alter the optical properties at the insulator-CO interface. In this sense, COs are good active materials for EO modulation. Indeed, many EO modulators have been proposed based on COs as the active material [12,13,14,15,16,17,18]. To apply a CO as active medium in an EO modulator, one challenge is that the field-effect-induced charge layer is only few nanometers thick. To enhance the impact of the charged layer on light absorption, three strategies are taken in this work.

First, the CO is integrated into an MIM plasmonic waveguide. Surface plasmons as hybrid electrical-optical waves give rise to strong confinement, which makes their propagation to be extremely sensitive to minor changes in the optical properties of the guiding materials. This provides a remarkable opportunity for optical modulation. Various plasmonic EO modulators have been reported since the 1980s [19,20,21,22,23], including three based on COs [24,25,26]. In particular, Dionne, et al. [23] reported an EO modulator integrating a metal-oxide-silicon junction into two silver films, where the modulating electric field can switch the waveguide between guiding and cut-off states. However, its large extinction ratio extremely relies on the junction uniformity in the light propagation direction. Two more recent works from another group reported microscale plasmonic phase modulator [27] and plasmonic intensity modulator [28]. In both works, MIM plasmonic waveguides are employed and nonlinear polymers work as the active media with modulator waveguide length shrunk to 10s of micrometers. Recent experimental works [24,25,26] report the integration of the MIC structure on the silicon waveguide, in a plasmonic waveguide, and below a glass prism, respectively. Modulation extinct ratio is reported 1 dB/μm in Ref. [24], and 2.71 dB/μm in Ref. [25]. In both cases, the dynamical operation of the modulators has not been reported. The result from Ref. [26] shows that the part of large modulation may only work at low speed. Whether the modulation can work at high speed is still a question.

Second, HfO$_2$ is employed as the insulator layer to enhance the field effect. HfO$_2$ has high DC dielectric constant, large dielectric strength, and is transparent for NIR light. Basically, the MIC structure functions as a parallel plate capacitor. Thus, a high-k insulator can induce large surface charge for a given gate voltage.

Third, two MIC structures are designed back-to-back to double the field-induced charge, simultaneously on top and bottom sides of CO, for the same gate voltage. The use of this double capacitor gating scheme in modulators can date back to dual-channel EO modulators proposed in 1970s or earlier [29]. It was recently used in a graphene modulator [30] and a proposed silicon slot-waveguide-based EO modulator [31]. The entire structure in this work is thus a metal-insulator-CO-insulator-metal (MICIM) structure, as shown in Fig. 1. The CO used in this work is ITO. Another advantage of using the double insulator layers is to decrease the MIM plasmonic waveguide attenuation, which sharply decreases with the increase of insulator thickness, especially at the nanoscale.

This work shows that the three strategies can greatly improve the modulator performance and result in an EA modulator with a propagation length only 800 nm. To our knowledge, this is the first experimental demonstration of an EO modulator with a nanoscale waveguide length. The EO modulators in previous works [23,24,25,27,28] involve waveguides with nanoscale width or thickness, but the waveguide length is still at the microscale.

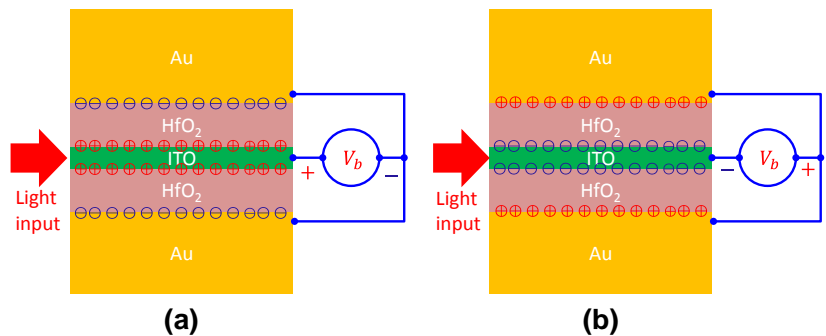

Figure 1. Illustration of the two working modes of the metal-insulator-CO-insulator-metal structure: (a) Depletion mode, where the ITO is less absorptive and the plasmonic waveguide has lower attenuation. (b) Accumulation mode, where the ITO is more absorptive and the plasmonic waveguide has higher attenuation.

As shown in Fig. 1, the principle of this EO modulator is quite straightforward. Light absorption in the gap between two gold films is controlled by the electric-field-induced charge in the ITO layer. Each MIC structure functions as a parallel plate capacitor. The field induced charge on each side of ITO can be calculated according to $Q = \varepsilon E A$, where $\varepsilon$ is the DC dielectric constant of HfO$_2$, $E$ is the applied electric field across the HfO$_2$ layer, and $A$ is the capacitor area. According to the bias polarity and strength, an MIC structure may work in two well-known modes. When a positive bias is applied at the ITO layer, depletion (of electrons) ($Q > 0$) occurs. Effectively, the thickness of ITO layer decreases and the MICIM waveguide becomes less absorptive, resulting low attenuation (i.e. ON state). When a negative bias is applied at the ITO layer, accumulation (of more electrons) ($Q < 0$) occurs. The ITO layer becomes more conductive and the MICIM waveguide becomes more absorptive, resulting high attenuation (i.e. OFF state). The third working mode of a metal-oxide-semiconductor structure, inversion (of carriers into holes), is less likely to occur.

Quantitatively, the optical dielectric constant $\epsilon$ of the ITO can be approximated by its free carrier concentration according to the Drude model,

$$\epsilon = \epsilon' + j\epsilon'' = \epsilon_\infty - \frac{\omega_p^2}{\omega(\omega + j\gamma)} \tag{1}$$

where $\epsilon_\infty$ is the high frequency dielectric constant, $\omega_p$ is the plasma frequency, and $\gamma$ is the electron damping factor. $\epsilon'$ and $\epsilon''$ are real and imaginary parts of optical dielectric constant, respectively. Note plasma frequency $\omega_p = \sqrt{\frac{Nq^2}{\epsilon_o m^*}}$, depending on carrier concentration $N$, and the effective electron mass $m^*$. $q$ represents the elementary charge. To make $\omega_p$ located in the NIR regime, the carrier concentration, $N$, should reduce to $10^{20}\sim10^{22}$ cm$^{-3}$, coinciding to that of COs. Outside of the range, the effect of same level of electric-field-induced charge becomes less significant on waveguide attenuation [13]. The specific value of carrier concentration can be controlled by the growth/deposition processes and post-annealing conditions [32].

The schematic of the device is illustrated in Fig. 2(a). The device fabrication consists of a series of layer-by-layer processes. For accurately patterning each layer, general photolithography and lift-off processes are applied. The fabrication starts from the deposition of 10 nm thick Ti adhesion and 100 nm thick gold layers on quartz substrate. Then, a 20-nm HfO$_2$ is deposited as the first buffer insulator by atomic layer deposition (ALD). The next process is the sputtering of 10-nm active material, ITO, at a substrate temperature of 60°C. After that, another 20 nm thick HfO$_2$ is deposited to provide the second buffer layer. Finally, 150-nm thick gold is deposited to define the light input port, the modulator length as well as to form electric contacts.

To test the device, the sample is flipped upside down as illustrated in Fig. 2(a). For easy description, the quartz substrate is treated as the top layer. As can be seen in Fig. 2 (b, c, d), there are three modulators with waveguide lengths, 3.0 μm, 800 nm, and 2.0 μm, respectively. The three modulators share the same GSG contact electrodes. Our investigation is focused on the 800 nm waveguide. Figures 2 (c & d) are the top and bottom side microscope images of our fabricated device. Figure 2(e) is the image of the edge of the device taken from the bottom side showing the multi-layer stacks.

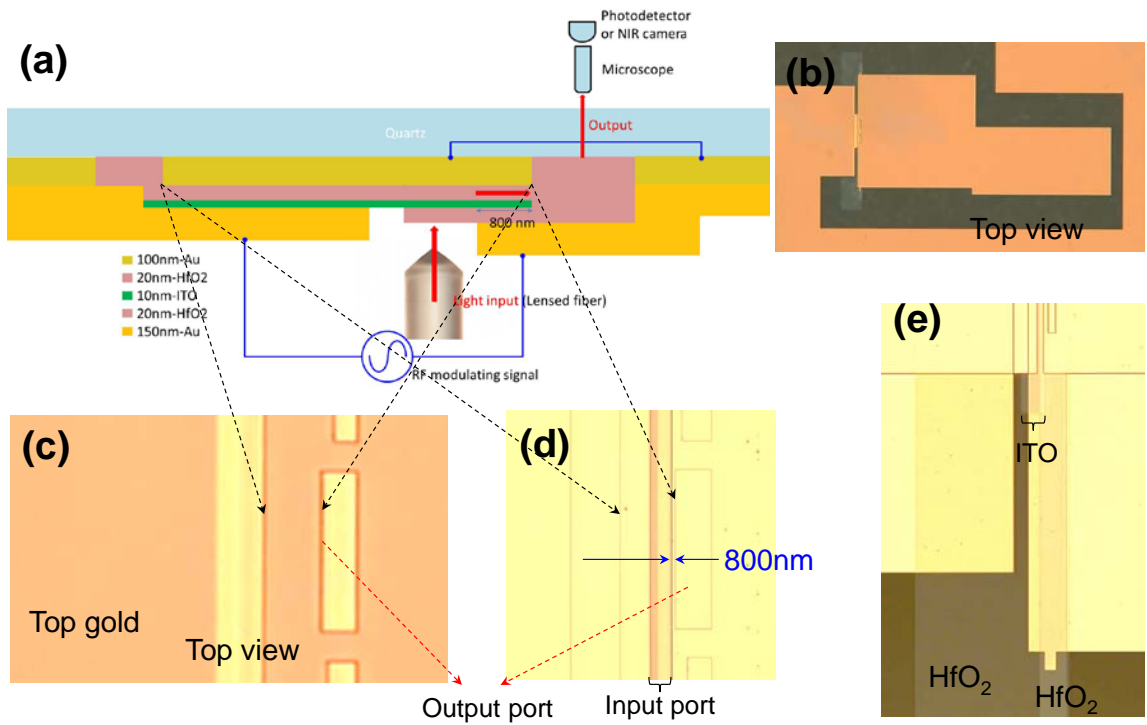

Figure 2. (a) Schematic of the fabricated EO modulator and the test setup. (b) Top view of the modulators and their GSG contact electrodes. (c) Top view of the output port. (c) Bottom view of the device input and output ports. (d) Bottom view of the edge of the device.

Light from a tunable laser is amplified by an erbium-doped fiber amplifier (EDFA) and then coupled into the sample through a lensed fiber from the bottom side of the sample. The lensed fiber can focus NIR light into a ~2 μm diameter spot at the working distance of ~15 μm. The entire sample is covered by either the top or bottom gold layer, or both layers. For easy explanation, the top and bottom gold layers are shown as slightly different colors. The gold layers are thick enough so that light cannot directly propagate through as shown in Fig. 2(b & c). The input port is a 10 μm wide slit in the bottom gold layer as shown in Fig. 2(d). Light propagation through the sample has to follow the gaps, i.e. plasmonic waveguides, between the two gold layers. See Fig. 2(a). Two MIM plasmonic waveguides simultaneously exist in our modulator. The left side one is a Au-HfO2-ITO-Au waveguide, which has very large attenuation and its length is designed to be 4 μm. Thus, the light transmission through the Au-$HfO_2$-ITO-Au waveguide is negligible. The right side one is a Au-$HfO_2$-ITO-$HfO_2$-Au waveguide, which is our modulating waveguide. It has much smaller attenuation, owing to the double $HfO_2$ layers, and much short length, only 800 nm, terminated by a slit in the top gold layer, which works as the output port. Therefore, the input and output ports are linked by the Au-$HfO_2$-ITO-$HfO_2$-Au waveguide.

As illustrated in Fig. 2(a), the top gold layer seems three separate parts, but the middle and right parts actually represent a continuous film with a 5μm-by-30μm rectangular window, as shown in Fig. 2(c), which means the top gold layer only has two electrically insulated parts. The bottom gold layer is also two insulated films: the left part works as a metal contact electrode for ITO; the right part simultaneously works as another electrode and a confining layer of the Au-$HfO_2$-ITO-$HfO_2$-Au plasmonic waveguide. As can be seen in Fig. 2(a), two confining gold layers of this waveguide are electrically connected at the right end. Consequently, when a voltage is applied between ITO and bottom gold layer, the same voltage is always applied between ITO and the top gold layer. The result is the double capacitor gating scheme as shown in Fig. 1. The gate voltage comes from a function generator and is applied to the electrodes through a microwave GSG probe from the sample bottom side. See Fig. 2(b).

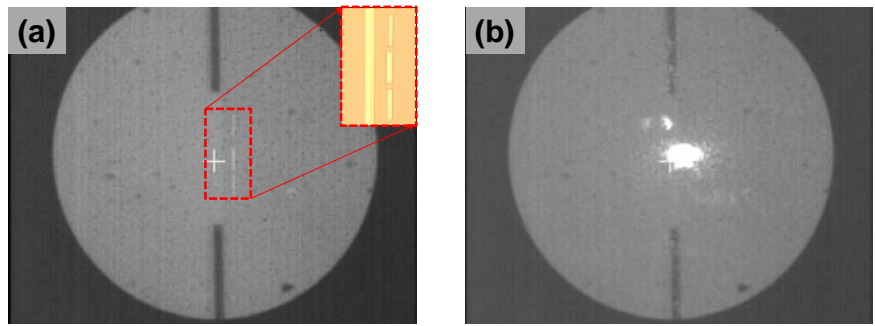

Figure 3. (a) Image of the output ports by the NIR camera. (b) Image of the output of the device through the NIR camera when NIR light is fed in from the input port.

In the experiment, a microscope, installed with an NIR camera at the top through an adapter, is placed over the sample for observation, alignment, and NIR imaging. The microscope is first adjusted to locate the output port. Figure 3(a) shows the image of the sample by the NIR camera. Then amplified NIR light (λ = 1560 nm) is fed into the input port through a lensed fiber. After the light propagates through the device, it is scattered into the quartz substrate and then into the air. Partial scattered light is imaged by the microscope NIR camera. Figure 3(b) shows the light scattering from the middle output window.

When an AC voltage (14 Vpp) is applied on the device, the change of power level of the output light with the voltage fluctuation can be observed. This verifies that the light output is really modulated. The modulation is so obvious that light intensity change in the NIR image can easily be seen by eyes at extremely low frequencies. Our supplemental movies document the modulation at $f$ = 3 Hz. Movies 1 and Movie 2 show the output from the modulators with waveguide lengths 800 nm and 2.0 μm, respectively.

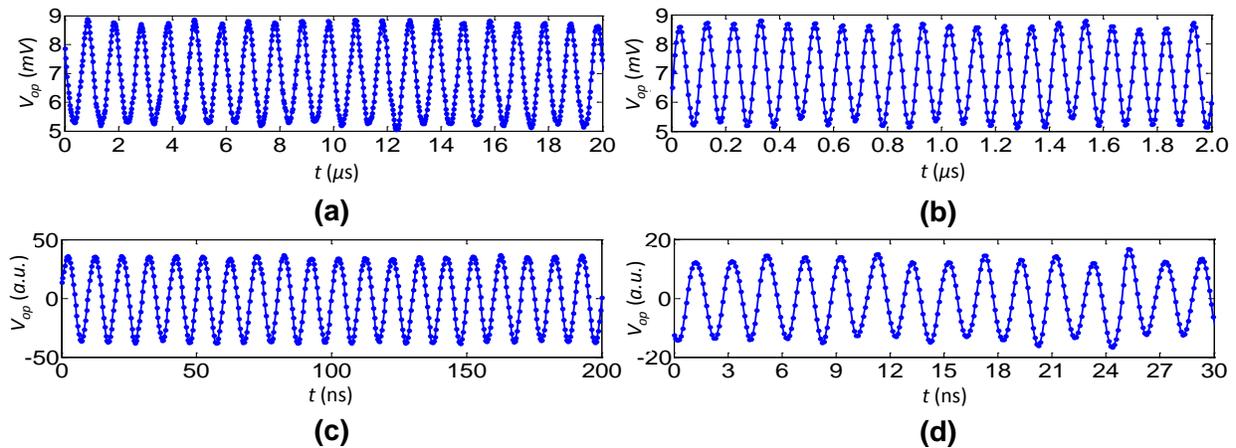

Figure 4. Oscilloscope measurement of the output light power under an applied electric signal of (a) 1 MHz, (b) 10 MHz, (c) 100 MHz, and (d) 500 MHz. (a&b) are obtained using a DC-coupled photodetector. (c&d) are obtained using an AC-coupled photodetector.

To test response of the device to higher frequency modulating voltage, the NIR camera is first replaced by a DC-coupled InGaAs photodetector (Thorlabs PDA10CF), which is positioned exactly at the focused image of the microscope adapter. The photodetector collects the output light power and converts it into an electric signal in an oscilloscope. A 50-Ω load is used to convert the photocurrent into voltage, $V_{op}$. Figures 4 (a&b) depict the tested photo voltage of the EA modulator under an applied 14

Vpp RF sine signal of 1 MHz and 10 MHz, respectively. If the extinction ratio is defined by $r_e = \frac{\max\{V_{op}\}}{\min\{V_{op}\}}$, the extinction ratio measured at 10 MHz is around 1.75 or 2.43 dB. To demonstrate the operation of the modulator at even higher frequency, the DC-coupled photodetector is replaced by an AC-coupled, amplified InGaAs photodetector (Newport 818-BB-30A), and the output is amplified by a microwave amplifier. Figures 4 (c&d) plot the tested photo voltage of the EA modulator under an applied 14 Vpp RF sine signal with frequencies $f$ = 100 MHz and $f$ = 500 MHz, respectively.

Owing to the limitation in the speed of the photodetector, the power level of the incident light, and the input coupling efficiency we observed light modulation only up to 500 MHz. The obtained results show a very promising prospective for an ultra-compact, ultra-fast optical modulator. We are upgrading our testing setup to evaluate its performance at GHz modulating frequencies. The results will be reported in the future.

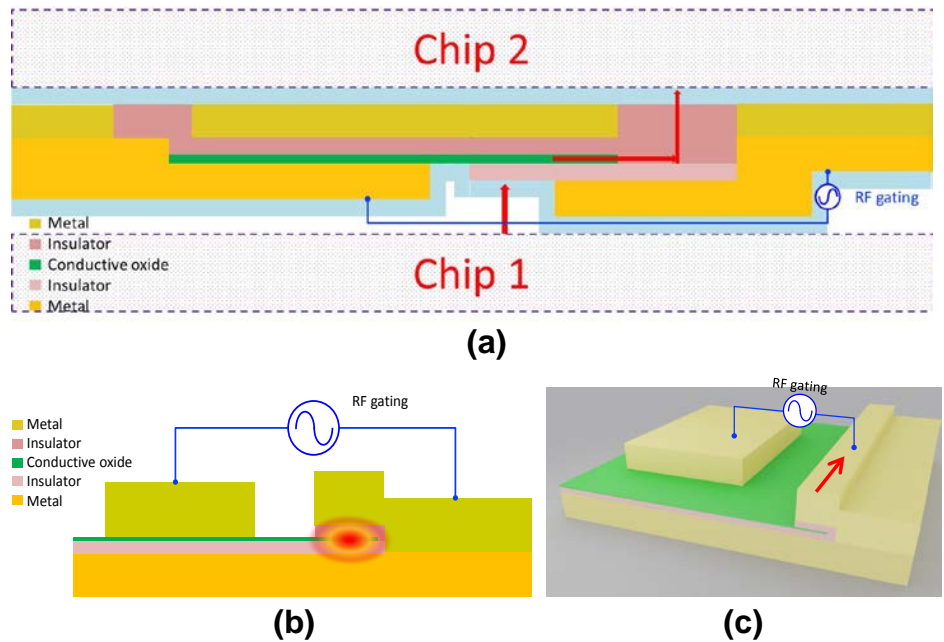

Figure 5. The potential applications of the MICIM modulator in 3D interconnection and in integrated photonic/plasmonic circuits. (a)  The potential application of the modulator in 3D interconnection, where both sides of the modulators are passivated and Chip 1 and Chip 2 may be standard CMOS circuits. (b, c) The potential application of the MICIM modulator in integrated photonic/plasmonic circuits, where (b) illustrates the front view, and (c), 3D view.  Photonic/plasmonic power flow is between the two metal layers in the red arrow direction.

In conclusion, we have experimentally demonstrated an 800-nm-long plasmonic EA modulator which has a significant modulation depth and can potentially work at GHz high speed at telecommunication wavelengths. In this device, we have combined the MIC structure with an MIM plasmonic waveguide configuration, with double capacitor gating scheme for greatly enhanced field effect. The involved fabrication processes are undemanding and CMOS-compatible. The device may meet the requirements for an ultra-compact and ultra-fast EO modulator as future 3D interconnection, where NIR light propagates between top and bottom sides of chip-based systems and is modulated by electric gate. To this end, both sides of the modulators are passivated and Chip 1 and Chip 2 may be standard CMOS circuits. As shown in Fig. 5, a variation of the modulator, yet based on the same principle, may also be employed for future on-chip nanophotonic/nanoplasmonic interconnection.

This material is based upon work supported by the National Science Foundation under Award No. ECCS-1308197. The authors would like to thank Peichuan Yin for his work on the 3D drawing and supplemental movies.

Contribution of authors: Z. Lu conceived the ideas, performed related calculation, and designed the modulators. K. Shi designed lithography masks, figured out the fabrication processes, and fabricated the modulators. K. Shi and Z. Lu tested the modulators, analyzed the experimental results, and prepared the manuscript.